\documentclass[aip, apl, reprint, amsmath, amssymb, superscriptaddress]{revtex4-2}

\usepackage[T1]{fontenc}
\usepackage{palatino}
\usepackage{soul}
\soulregister\cite7
\soulregister\ref7
\soulregister\pageref7
\soulregister\ce7
\usepackage{lipsum}
\soulregister\ref{7}
\usepackage{graphicx}
\usepackage[svgnames]{xcolor}
\usepackage{amsmath, bm}
\usepackage[detect-none]{siunitx}
\usepackage{fancyref}
\usepackage[colorlinks,
            linkcolor=Blue,
            urlcolor=Blue,
            citecolor=DarkGreen]{hyperref}
\usepackage[utf8]{inputenc}
\usepackage{textcomp}
\usepackage[version=3]{mhchem}
\usepackage{miller}
\usepackage{tabularx}
\usepackage{pdfpages}
\usepackage{pgffor}
\usepackage{etoolbox}
\makeatletter
\patchcmd{\@outputpage@head}{\@ifx{\LS@rot\@undefined}{}{\LS@rot}}{}{}{}
\makeatother

\begin{document}

\title[]{The role of etching anisotropy in the fabrication of freestanding oxide microstructures on \ce{SrTiO3(100)}, \ce{SrTiO3(110)}, and \ce{SrTiO3(111)} substrates}

\author{Alejandro Enrique \surname{Plaza}}
\affiliation{CNR-SPIN, C.so F.\,M.~Perrone, 24, 16152 Genova, Italy}

\author{Nicola \surname{Manca}}
\email{nicola.manca@spin.cnr.it}
\affiliation{CNR-SPIN, C.so F.\,M.~Perrone, 24, 16152 Genova, Italy}

\author{Cristina \surname{Bernini}}
\affiliation{CNR-SPIN, C.so F.\,M.~Perrone, 24, 16152 Genova, Italy}

\author{Daniele \surname{Marré}}
\affiliation{Dipartimento di Fisica, Università degli Studi di Genova, 16146 Genova, Italy}
\affiliation{CNR-SPIN, C.so F.\,M.~Perrone, 24, 16152 Genova, Italy}

\author{Luca \surname{Pellegrino}}
\affiliation{CNR-SPIN, C.so F.\,M.~Perrone, 24, 16152 Genova, Italy}

% \date{\today}

\begin{abstract}
  The release process for the fabrication of freestanding oxide
  microstructures relies on appropriate, controllable and repeatable
  wet etching procedures. \ce{SrTiO3} is among the most employed
  substrates for oxide thin films growth and can be decomposed in
  HF:water solution. Such process is strongly anisotropic and is
  affected by local defects and substrate cut-plane. We analyze the
  etching behavior of \ce{SrTiO3} substrates having (100), (110), and
  (111) cut-planes during immersion in a 5\% HF:water solution. The
  etching process over the three substrates is compared in terms of
  pitting, anisotropy, macroscopic etch rate and underetching effects
  around HF-resistant \ce{(La{,}Sr)MnO3} thin film micropatterns. The
  release of targeted structures, such as the reported
  \ce{(La{,}Sr)MnO3} freestanding microbridges, depends on the
  substrate crystallographic symmetry and on the in-plane orientation
  of the structures themselves along the planar directions. By
  comparing the etching evolution at two different length scales, we
  distinguish two regimes for the propagation of the etching front: an
  intrinsic one, owning to a specific lattice direction, and a
  macroscopic one, resulting from the mixing of different etching
  fronts. We report the morphologies of the etched \ce{SrTiO3}
  surfaces and the geometries of the underetched regions as well as of
  the microbridge clamping zones. The reported analysis will enable
  the design of complex MEMS devices by allowing to model the
  evolution of the etching process required for the release of
  arbitrary structures made of oxide thin films deposited on top of
  STO.
  \\
  \\
  This article may be downloaded for personal use only. Any other use
  requires prior permission of the author and AIP Publishing. This
  article appeared in ``A. Plaza \textit{et al.},
  Appl. Phys. Lett. \textbf{119}, 033504,  (2021)'' and may be found at
  \href{http://dx.doi.org/10.1063/5.0056524}{http://dx.doi.org/10.1063/5.0056524}

\end{abstract}

\maketitle

\ce{SrTiO3} (STO) is a standard substrate for the deposition of many
transition metal oxiedes (TMO) thin films\cite{Schlom2008} and for the
realization of oxides-based devices due to its convenient lattice
parameter, variety of surface preparation procedures, wide band gap,
and large dielectric constant\cite{Biswas2017} and, recently, it has
been employed as preferential substrate for developing new oxide MEMS
devices\cite{Pellegrino2009, Biasotti2010, Deneke2011, Manca2013,
  Ceriale2014, Manca2015, Manca2019a} without requiring integration
procedures with Si technology.\cite{Baek2011, LeBourdais2015,
  Bhaskar2016, Liu2019a, Lim2020, Nascimento2021} Common commercially
available \ce{SrTiO3} substrate crystal cuts are (100), (110) and
(111). STO(100) is the most studied and used one, it is weakly polar
and has the highest surface symmetry. In contrast, STO(110) and
STO(111) crystal cuts are polar and have a reduced in-plane symmetry
that makes them prone to surface reconstructions and adsorbates that
influence device fabrication steps.\cite{Tasker1979, Sanchez2014,
  Chang2008} The proposed etching model of single-crystal \ce{SrTiO3}
is the so-called ``Dislocation hierarchical tree
structure'',\cite{Szot2018} which takes into account both rate
dispersion and induced surface morphology and many critical details of
this behavior have been tested so far.\cite{Wang1998, Jin2013,
  Waser2009} A complementary concept to this crystallographic approach
is the ``Reactive surface area'', associated to methods suitable to
study the variability of etch rates and mechanisms on a bigger
scale.\cite{TrindadePedrosa2019, Luttge2019, Casey1988}
Notwithstanding the significative overall results briefly reviewed,
\ce{SrTiO3} specific results regarding crystal cut, anisotropy and
etch rate variability in the case of MEMS fabrication are still
lacking. In such a case, many parameters are fixed due to device
specification constraints. Hence, the specific effects of crystal cut
and device orientation, are very relevant.In order to develop a MEMS
technology based on crystalline oxide thin films deposited on top of
\ce{SrTiO3} substrates, it is thus crucial to understand the evolution
of the substrate etching process along different lattice directions,
in particular with respect to the planned device geometry and feature
size.

In this work, we analyze faceting, etch rates, pitting and
underetching effects of chemically etched (5\% HF aqueous solution)
STO substrates having different cut-planes, namely (100), (110), and
(111). We employ 5$\times$5$\times$0.5\,mm${}^3$ single crystal
substrates from CrysTec GmbH with miscut angle below 0.1$^\circ$. We
compare the etching evolution on two different device geometries
designed to measure both the STO etching rates at the macroscale and
the etching dynamics at the microscale. For the former, we use a hard
mask made from a 50\,nm thick \ce{(La_{0.7}, Sr_{0.3})MnO3} (LSMO)
film with 200$\times$200\,\textmu m${}^2$ square holes array. For the
latter, we investigate the under-etching and the release process of
100\,nm thick and 5\,µm wide LSMO microbridges.

\begin{figure*}[]
 	\includegraphics[]{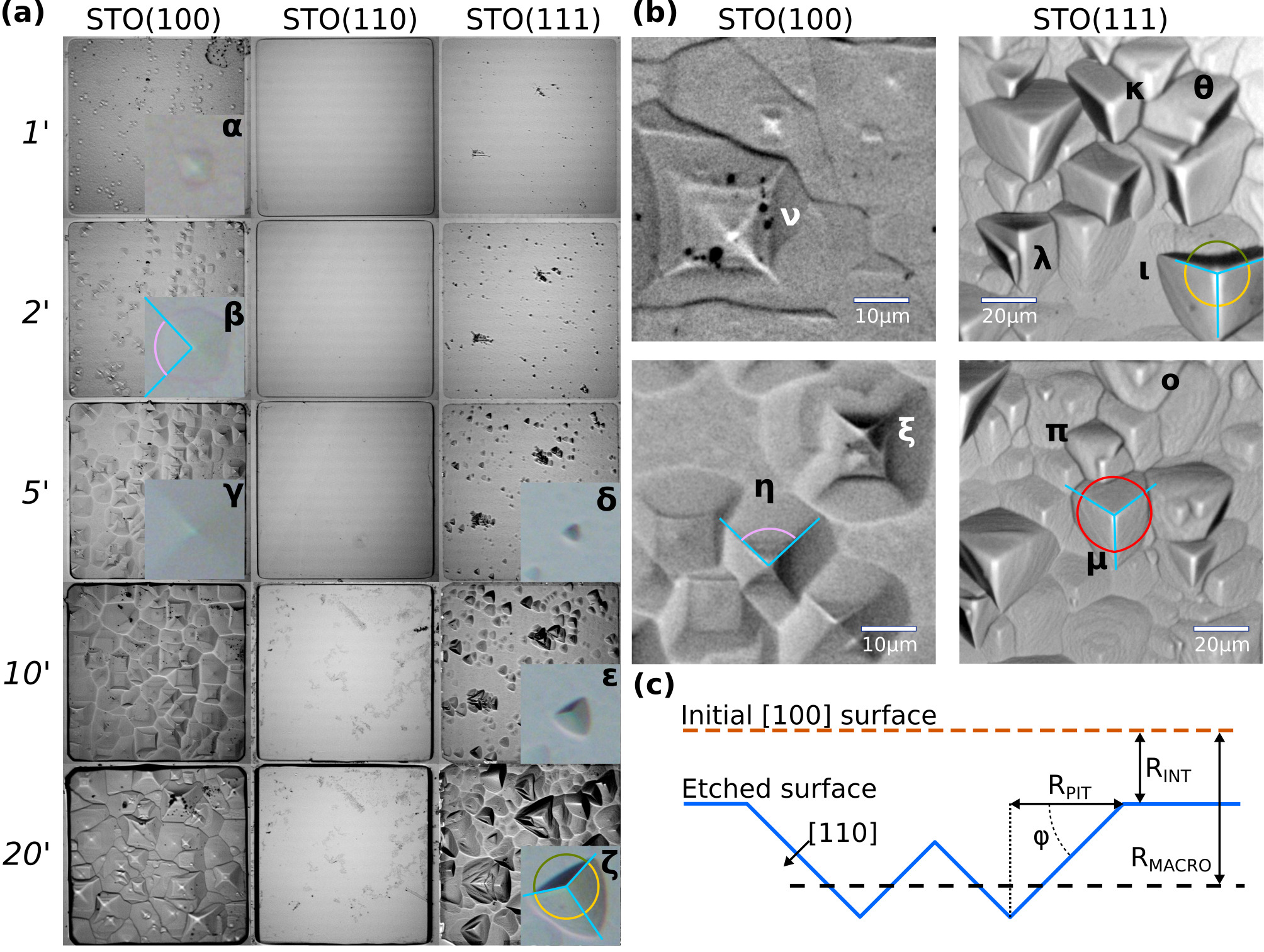}
 	\caption{\label{fig1} Optical microscopy images of etched
          square holes of 200\,\textmu m lateral size.  (a) Pictures
          taken at different etching times. Insets are magnified pit
          details. The visible dark-grey marks on the STO(110) surface
          are not pits, but residues of strontium fluoride coming from
          the STO chemical etching process (see Supplementary
          Material, Sec.~I).  (b) Details of the etched surface, where
          the STO(100) pictures were acquired after 5 minutes of
          etching while the STO(111) ones after 60 minutes. In-plane
          angles are 90$^\circ$ (pink), 110$^\circ$ (green),
          140$^\circ$ (yellow), 120$^\circ$ (red).  (c) Simplified
          illustration of the etching evolution in the out-of-plane
          direction for a STO(100) substrate.}
\end{figure*}

LSMO films are grown by Pulsed Laser Deposition (PLD) from a sintered
powder target using an excimer laser (248\,nm wavelength) having beam
energy density of 0.6\,J/cm${}^2$ and 2\,Hz repetition rate. The
deposition temperature is 800\,$^\circ$C and the oxygen background
pressure 10${}^{-4}$ mbar. After the growth, the films are in-situ
annealed for 20 minutes at 600\,$^\circ$C and 200\,mbar of oxygen
pressure. Both hard mask and microbridge samples are fabricated by
standard optical lithography as described in previous works from our
group.\cite{Pellegrino2009, Ceriale2014} The hard mask etched samples
are obtained by put soaking together (100), (110) and (111) STO
samples in HF (5\% in aqueous solution) kept at 30\,$^\circ$C.  This
allows us to perform some statistical analysis and test for
homogeneity over the sample surface. At given times (1, 2, 5, 10, 20,
40 and 60 minutes), they are all removed from the bath, cleaned in
deionized water, dried under nitrogen flow, and inspected at the
optical microscope.  LSMO microbridges oriented along three in-plane
orientations (0$^\circ$, 45$^\circ$, and 90$^\circ$) were realized on
three substrates having different cut-planes and fabricated following
the same procedure discussed above, with the only difference that
after 60 minutes of etching they were dried using a Critical Point
Drying system.  HF decomposes the STO, leaving water-insoluble salts
on the surfaces that are identified as strontium fluoride (see
Supplementary Material, Sec.~I). The removal of such salts is achieved
mechanically during the etching process by magnetic stirring at the
relatively slow rotational speed of $\approx$200 r.p.m., so as to
avoid the break of the microstructures by drag forces.

\begin{table*}[]
  \centering
  \begin{tabular}{c l r r r c l}
    \hline\hline
    Direction& Method            &Rate eq.            ~&~ RMS                & Etchant & Temperature & Reference\\
    \hline
    [001]    & R$_\mathrm{MACRO}$ &13.5 \,\textmu m/h ~&~ 1\,\textmu m    & 5\% (2.8M) HF  & 30$^\circ$C & This work\\\relax
    [011]    & R$_\mathrm{MACRO}$ &9 \,\textmu m/h    ~&~ 4\,nm           & 5\% (2.8M) HF  & 30$^\circ$C & This work\\\relax
    [111]    & R$_\mathrm{MACRO}$ &0.6 \,\textmu m/h  ~&~ 0.75\,\textmu m & 5\% (2.8M) HF  & 30$^\circ$C & This work\\\relax
    [001]    & Surface           &1.2 \,\textmu m/h  ~&~ N/A             & 2.5M HF & 30$^\circ$C & Ref.~\onlinecite{Spalding1999}\\\relax
    [001]    & Pits depth        &11.1 \,\textmu m/h ~&~ N/A             & 2.5M HF & 30$^\circ$C & Ref.~\onlinecite{Spalding1999}\\\relax
    [001]    & Pits depth        &21.6 \,\textmu m/h ~&~ N/A             & N/A     & N/A         & Calculated from Ref.~\onlinecite{Szot2018}\\\relax
    [001]    & Terraces          &0.01 \,\textmu m/h ~&~ N/A             & Buffered HF     & ``Room Temperature'' & Calcualted from Ref.~\onlinecite{Lippmaa1998}\\\relax
    [011]    & Surface           &6.0 \,\textmu m/h  ~&~ N/A             & 2.5M HF & 30$^\circ$C & Ref.~\onlinecite{Spalding1999}\\
    \hline
    [001]    & R$_\mathrm{PIT}$   &5 \,\textmu m/min  ~&~ N/A             & 5\% (2.8M) HF  & 30$^\circ$C & This work\\\relax
    [111]    & R$_\mathrm{PIT}$   &1.3 \,\textmu m/min~&~ N/A             & 5\% (2.8M) HF  & 30$^\circ$C & This work\\
    \hline\hline
  \end{tabular}
  \caption{Comparison etching rate for \ce{SrTiO3} single crystal in
    HF bath along the [001], [011], and [111] directions. In our work
    the etch front rate is calculated by considering the final depth
    after 60\,mins. The RMS evaluates the surface roughness and is
    calculated over the entire 200$\times$200\,\textmu m$^2$ square
    mask (see also the Supplementary Material, Sec~II and III). For an
    in-depth comparison a particular care with respect to how the rate
    is defined in each of the referenced works is required.}
\label{tab1}
\end{table*}

We start by discussing the etching process at the macroscale by
monitoring two characteristics of the STO substrate regions that are
not covered by the LSMO hard mask: (1) the formation, alignment, and
evolution of pyramidal etch pits; (2) the overall substrate etching in
the out-of-plane direction.

The pits form and progressively grow on the STO(100) and STO(111)
substrates, resulting in surfaces covered with well-formed pyramidal
pits of quite uniform lateral dimension after 10 min and 40 min for
STO(100) and STO(111), respectively (see Figure~\ref{fig1}(a)).
STO(100) pits have square symmetry with a centered apex as shown in
Fig.~\ref{fig1}(a) ($\alpha$, $\beta$, $\gamma$) and (b)($\eta$),
while STO(111) pits have triangular section, with the apex mostly
shifted from the symmetry center, as visible from the shadowed
bounding plane in Fig.~\ref{fig1}(b) ($\theta$, $\iota$, $\kappa$,
$\lambda$) and in some case with a remarkable symmetry, as in
Fig.~\ref{fig1}(b) ($\mu$).  For both STO(100) and STO(111)
substrates, the pits bounding planes are parallel to (hh0) substrate
planes and all the pits are identically oriented with respect to the
substrate.  As the etching proceeds, already existing pits grow in
size, but nucleation of new pits is not observed over the pristine STO
surface.  During pit growth, other localized events can be found that
are able to change pit bounding planes direction, as in
Fig.~\ref{fig1}(b) ($\lambda$), or even the pit morphology (i.e. a
pyramidal pit within a flat pit), as in Fig.~\ref{fig1}(b) ($\nu$,
$\eta$, $\pi$, $o$).  By following the evolution of individual pits,
as shown in Fig.~\ref{fig1}(a) ($\alpha$, $\beta$, $\gamma$) and
($\delta$, $\varepsilon$, $\zeta$), it is possible to estimate their
in-plane growth rate (R${}_\mathrm{PIT}$ of Fig.~\ref{fig1}(c)),
reported in Table~\ref{tab1}, as long as they do not coalescence. The
indicated growth rates are in agreement with literature
\cite{Spalding1999}.  Finally, once the surface is cluttered with
overlapping pits, the etching proceeds distributed all over the
exposed substrate surface.  In contrast, the STO(110) substrate
remains pretty flat even at the end of the etching process, with a
r.m.s. roughness of about 4\,nm, as measured by AFM, and no detectable
pit formation (see Supplementary Material, Sec.~II). Such striking
difference shows, in accordance with previous literature reports, that
the [110] direction is a fast etching direction, since no pit
eventually nucleated on this face would be able to survive on its
own.\cite{Spalding1999}

As schematically illustrated in Fig.~\ref{fig1}(c) (showing the
STO(100) case), the overall out-of-plane etching rate on a large scale
(several tens of micrometers) is a combination of different processes,
also including the evolution of the pits. In fact, the etch front
along a direction (intrinsic rate R$_\mathrm{INT}$) given by a
crystallographic face may proceed by the removal of crystalline layers
in the same crystallographic direction or by removing layers in
different and faster etching directions, enhancing the overall etching
rate through the formation of pits (in plane pit growth rate
R${}_\mathrm{PIT}$). Moreover, during the pitting process, a flat
surface along a slow etching direction may form (i.e. truncated
pyramidal pits) decreasing the etching speed unless new pits are
formed. Considering these mechanisms, the nucleation and the growth of
different etching planes may depend on local conditions (i.e. etchant
concentration gradients, local fluid velocity, salt precipitation and
contamination, etc.). However, taking into account that the observed
pit characteristic length is well below the size of the exposed
regions in our hard mask, we can evaluate the average large-area
etching (R${}_\mathrm{MACRO}$) rate for STO(100), STO(110) and
STO(111) substrates by inspecting our samples after 60 minutes of
etching. This was performed by an optical profilometer (see
Supplementary Material, Sec.~III), and the results of this analysis
are reported in Table~\ref{tab1}, together with a comparison with
etching rates reported in previous works.

A second important aspect, directly related to the release process of
suspended microstructure, is the evolution of the underetch profile,
i.e.\ the STO region etched below the mask at the edges of patterned
structures. We now analyze the evolution of the underetching below the
LSMO microbridge and the geometry of the underetch profile around the
LSMO microbridge clamping regions for different substrate types and
microbridge directions.
\begin{figure}[]
	\includegraphics[width=\linewidth]{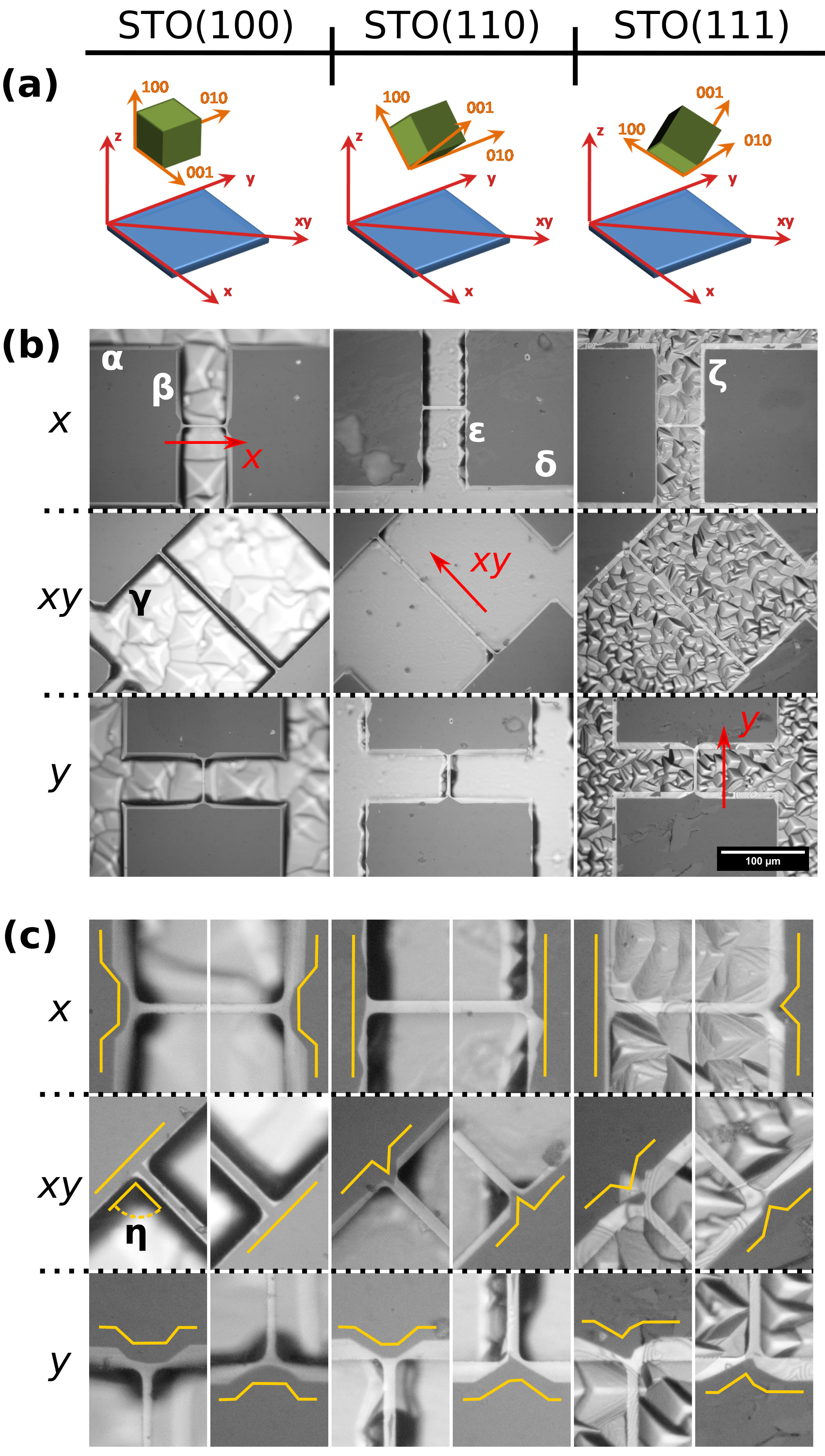}
	\caption{\label{fig3} (a) Scheme of the experiment, indicating
          the three STO samples and crystallographic directions: STO
          substrate surface (blue); sample directions (red); crystal
          unit cell (red); crystal basis directions (orange).  (b)
          Optical microscopy at 40X magnification after 40 minutes
          etching time of LSMO microbridges of different orientations
          (rows) on STO substrates with different crystal cut
          (columns).  (c) Magnified detail of the clamping zone of
          microbridges (Profile sketch in red). Light grey regions are
          freestanding.}
\end{figure}

\begin{figure}[]
  \includegraphics[width=\linewidth]{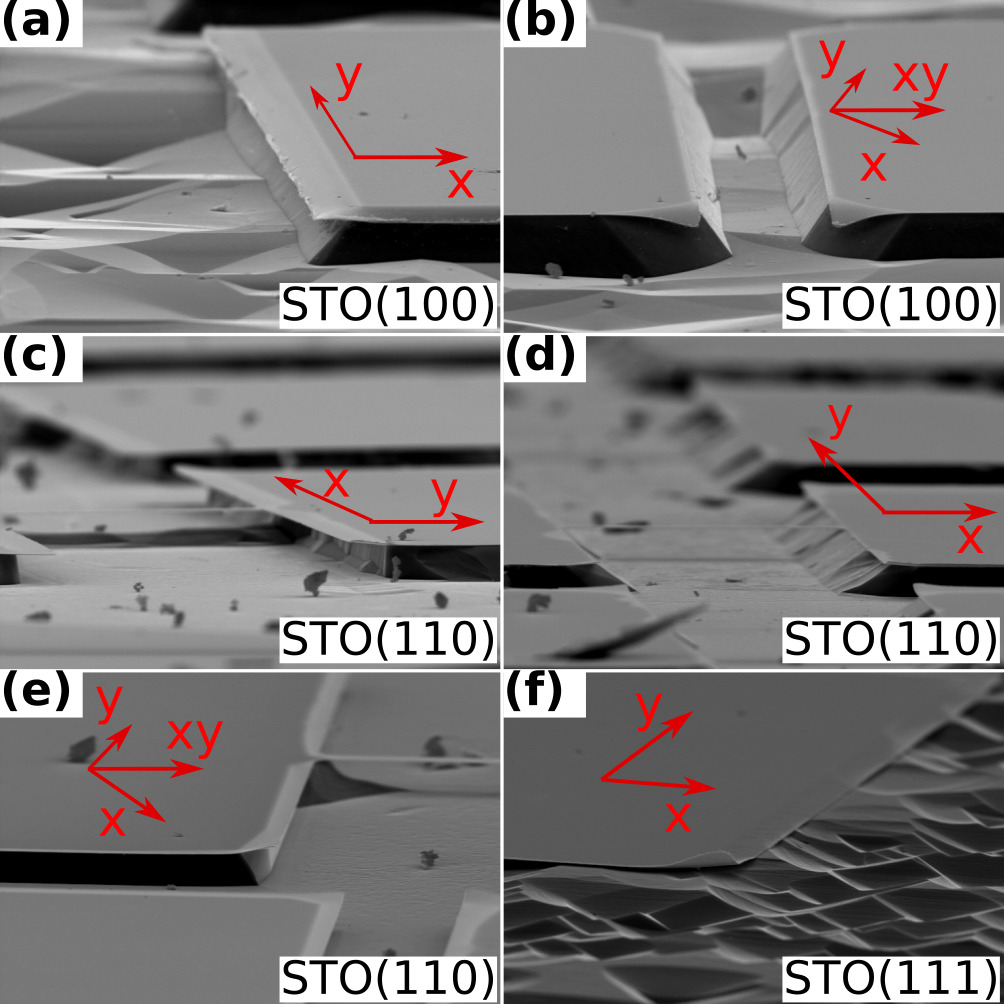}
  \caption{\label{fig4} SEM images of LSMO microbridges between square
    pads. In-plane directions are reported in red color.}
\end{figure}

In Figure~\ref{fig3} and Figure~\ref{fig4} suspended regions appear as
light-gray colored and progressively evolve around all the borders of
the LSMO patterns.  The optical inspection (Fig.~\ref{fig3}) monitors
the 2D projection of the underetch profiles only, while their
tridimensional structure is better evidenced in the SEM pictures
reported in Fig.~\ref{fig4}. In the following, in-plane directions are
defined as ‘$x$’, ‘$y$’ and ‘$xy$’, corresponding to the substrate
edges and diagonal, respectively, while ‘$z$’ is the out-of-plane
direction, as indicated in Fig.~\ref{fig3}(a). In the STO(100) the
extension of the underetched region is similar in both the equivalent
directions $x$ and $y$ (Fig.~\ref{fig3}(b) ($\alpha$, $beta$)), but
smaller for bridges oriented along the $xy$ direction
(Fig.~\ref{fig3}(b) ($\gamma$)). STO(110), instead, presents a less
symmetric pattern with a straight underetch profile in the $y$
direction, characterized by vertical walls (Fig.~\ref{fig3}(b)
($\delta$) and Fig.~\ref{fig4}(c). This is in contrast with the rough
underetch profile in $x$ and $xy$ directions (Fig.~\ref{fig3}(b)
($\varepsilon$), and Fig.~\ref{fig4}(d). For the STO(111) case, the
underetch profile is characterized by a larger extension below the
LSMO pattern and a marked asymmetry between the $+x$ and $-x$
directions, as evidenced in Fig.~\ref{fig3}(b) ($\zeta$). The etching
in the $z$ direction is limited and characterized by a faceted
surface, as already discussed and visible in Fig.~\ref{fig4}(f). The
faceted STO underetch profile have different orientations with respect
to the bounding planes observed on pits forming in open regions of the
substrates (Fig.~\ref{fig4}); this fact is a clear indication that
geometric boundary conditions significantly contribute in determining
the final clamping profile of the etched structures.

The time required to obtain a complete release from the substrate
depends on the device geometry and size. As a general reference, we
can consider the reported structures, i.e.\ 5\,\textmu m-wide LSMO
microbridges, etching time to be about 30 minutes for all the cases
except the two combinations: STO(100) $xy$-aligned and STO(110)
$y$-aligned. In such cases, etching times exceeding 1\,h could be
required depending on local defects. Considering these etching times,
in general, the underetch profile is not the equilibrium one, as
predicted by the Wulff--Jaccodine construction.\cite{Jaccodine1962}
Instead, the profile is determined by the system anisotropy and
varies, among other factors, with the in-plane orientation of the
pattern, crystal cut-plane and etching time.
      
Regarding the etching at the clamping zones, the process proceeds even
after the microbridge is suspended. The precise geometry of the
clamping regions and the symmetry between the two edges of the bridges
is discussed in Fig.~\ref{fig3}(c). They depend on microbridge
orientation and substrate crystal cut as follows:
\begin{itemize}
\item STO(100): the clamping region is symmetric for all the three
  microbridge orientations $x$, $y$ and $xy$. In the case of the $x$
  and $y$-oriented microbridges, the clamping zone has a polygonal
  shape, while in those oriented in the $xy$ direction it has a net
  flat shape that progressively evolves towards a complete release of
  the microbridge.
\item STO(110): the geometry of the clamping region, although
  symmetric in both cases, differs significantly between the $x$ and
  $y$ directions. For $x$-aligned microbridges, it is flat and aligned
  with the borders of the pads, while for $y$-aligned microbridges it
  protrudes from the pads, outlining a polygon. This is not the case
  for STO(110) $xy$-oriented micro-bridges, where asymmetry arises.
\item STO(111): a varied mix of shapes and symmetries at the clamping
  regions can be observed. Microbridges aligned in the $x$ direction
  have symmetric clamping regions, but one with a flat profile and the
  other with protruding profile. Microbridges aligned along the y and
  xy directions have asymmetric clamping regions, with the former
  showing larger under-etch.
\end{itemize}

In the following, we show that the complex faceting observed in the
clamping regions results from the interplay between substrate
cut-planes, device geometry, and anisotropic etching. In addition, at
small scales the etching rate measured along different cut planes
(nearby pattern features) is different from what extracted from the
macroscopic rates R${}_\mathrm{MACRO}$ measured reported in
Table~\ref{tab1}, where the averaging effects over different
crystallographic directions plays a major role.  This also indicates
that the relevant length scale determining the geometry of the edges
in micrometric devices is below the threshold required to observe the
large-area rates.

The symmetric underetching behavior for STO(100) and the similarity
between the $x$ and $y$ directions reflects the crystallographic
equivalence between [001] and [010] directions. The lower underetch
along xy corresponding to the [011] direction, can be explained
considering that this is a fast etching direction, the xy oriented
microbridges progressively exhibit walls parallel to the [1hh]
directions that have a slower etching rate than the pure [011] and
[001] ones. The trapezoidal-shaped (in-plane) clamping zones is the
result of the [111] face etching that is slower than the [110] face
one, hence prevailing the concave corner.\cite{Jaccodine1962} The
90$^\circ$ angle between the underetch profiles (fig.~\ref{fig3}(c)
($\eta$)) observed in the clamped regions for the microbridges aligned
along the $xy$ direction can be explained considering the formation of
underetch faces oriented along equivalent [111] directions. The
profile symmetry is guaranteed by the crystallographic equivalence
between the involved directions on both sides of the same clamping
zone.

The STO(110) underetch in-plane asymmetry arises from the fact that
the $x$ and $y$ directions are aligned along the slow [100] and the
fast [110] etching directions, respectively. The flat clamping zone of
x-aligned microbridges is the result of reaching a steady state
condition with the exposure of the (100) faces, which have a slower
etching rate. The faceting of the profile observed around the clamping
regions along the $y$ direction results from the faceting vicinal to
the [110] direction, which determines deep trenches into the substrate
along the $y$-directed border regions. The clamping regions of
$y$-aligned microbridges have a completely different shape, because in
this case the underetch vertical walls are the steady state
slow-etching faces ([100] equivalent directions). In this case, other
residual crystallographic planes are exposed, such as the [111]. The
symmetry of the clamping regions in both $x$ and $y$ aligned
microbridges is due to the fact that, contrary to $xy$-aligned
microbridges, sagittal planes containing microbridges axis are
symmetry planes of the crystal lattice.

STO(111) has in-plane crystallographic trigonal symmetry that does not
match the symmetry of the microbridge patterns, leading to a variety
of clamping zone shape. Due to the lower etch rate in the $z$
direction and the relevant in-plane underetch, it is difficult to
measure the extension of the underetch profile. However, by applying
the same analysis used for both STO(100) and STO(110), we may argue
that microbridges aligned along the $x$ direction have one flat
clamping zone probably limited by the slow etching [100] face, while
the other clamping region, with a protruding profile, is limited by
the equivalent [010] and [001] directions. The $y$-aligned
microbridges clamping regions have asymmetric profiles and are
probably limited by the slow etching rate along the [010] and [001]
directions in one side and by the fast etching rate along the [110]
and [101] on the other side of the microbridge. In the case of
$xy$-aligned microbridges, analogous analysis leads to infer that the
clamping regions are limited by slow etching rate along the [010] and
[001] directions, each one at different angle with the microbridge
sagittal plane, resulting in an asymmetrical profile.

Finally, in order to demonstrate the design opportunity enabled by our
study, we show in the Supplementary Material Sec.~IV the realization
of a LSMO micro-mechanical bridge resonator that can be measured
optically by focusing a laser through a double-polished
\ce{SrTiO3(110)} substrate. This was made possible thanks to the low
RMS of the bottom surface, which is preserved after the chemical
etching, and the transparency of \ce{SrTiO3} to visible light. Such
configurations will allow to couple TMO-based resonators in proximity
to other systems while maintaining an optical readout scheme.

In conclusion, we showed that the choice of the STO crystal cut-plane
has significant effects in terms of etching rate, pitting of large
areas, microbridge release time, underetching and shape of the
clamping regions. With the shrinking of the device size, the presence
of pits and their size become increasingly important and, for
increasing complexity of MEMS geometries, achieving desired morphology
of the released structures will require trade-off solutions based on
choosing the proper crystal cut. Lattice defects play also a crucial
role in the propagation of the etching front and their statistical
spatial distribution determines a critical scale separating large and
small-area behavior of the etching process. Moreover, the in-plane
orientation of the microbridges has considerable implications on the
final geometry of the clamping regions, likely affecting its
mechanical behavior.  Below we provide few general considerations
summarizing the behavior of STO substrates having different cut-planes
with respect to the investigated chemical etching process. As a
general reference, the lattice symmetry of the substrate is the main
characteristic determining how freestanding regions are released and
this should be taken into account for oxide MEMS design.
\begin{itemize}
\item STO(100) shows 90$^\circ$ in-plane symmetry and the fastest
  macroscopic etching direction along the out-of-plane direction and
  good in-plane etching rates.
\item STO(110) shows smooth etched substrate surfaces and good
  out-of-plane etching rates. It also shows the possibility to obtain
  sharp underetch walls, taking into account for the device geometry
  or symmetry.
\item STO(111) has the fastest in-plane underetching rate and the
  lower out-of-plane etching rate
\end{itemize}

\section*{Supplementary Material}

Supplementary material includes the following: SEM image of the
deposits of salts scattered over the \ce{SrTiO3} surface and Energy
Dispersive X-ray (EDX) spectra; roughness analysis of etched STO(110);
optical profilometry images over large-area etched regions; mechanical
measurement of a LSMO MEMS through the STO(110) substrate.

\section*{Acknowledgments}

We thank Flavio Gatti, Francesco Buatier de Mongeot, and Lorenzo
Ferrari Barusso for providing access to the optical profilometer. This
work was carried out under the OXiNEMS project (www.oxinems.eu). This
project has received funding from the European Union’s Horizon 2020
research and innovation programme under grant agreement No 828784.

\section*{Data Availability Statement}

The data that support the findings of this study are openly available
in Zenodo at
~\href{http://dx.doi.org/10.5281/zenodo.4738478}{http://dx.doi.org/10.5281/zenodo.4738478}.

\bibliography{Library.bib}
% \bibliographystyle{apsrev4-1}

%This adds the supplementary at the end (for the ArXiv)

\newpage\newpage


\section*{Supplementary material}

*Supplementary Material*

—

# The role of etching anisotropy in the fabrication of freestanding oxide microstructures on SrTiO$_3$(100), SrTiO$_3$(110), and SrTiO$_3$(111) substrates

Alejandro Plaza,[1] Nicola Manca,[1, *] Cristina Bernini,[1]

Daniele Marré,[2, 1] and Luca Pellegrino[1]

[1]*CNR-SPIN, C.so F. M. Perrone, 24, 16152 Genova, Italy*

[2]*Dipartimento di Fisica, Università degli Studi di Genova, 16146 Genova, Italy*

This supplemental material contains the following:

- Section I: SEM image of the deposits of salts scattered over the SrTiO$_3$ surface and Energy Dispersive X-ray (EDX) spectra.

- Section II: Roughness analysis of etched STO(110)

- Section III: Optical profilometry images over large-area etched regions

- Section IV: Mechanical measurement of a LSMO MEMS through the STO(110) substrate

———

* nicola.manca@spin.cnr.it

## Sec. I. SEM IMAGE OF THE DEPOSITS OF SALTS SCATTERED OVER THE STO SURFACE AND ENERGY DISPERSIVE X-RAY (EDX) SPECTRA

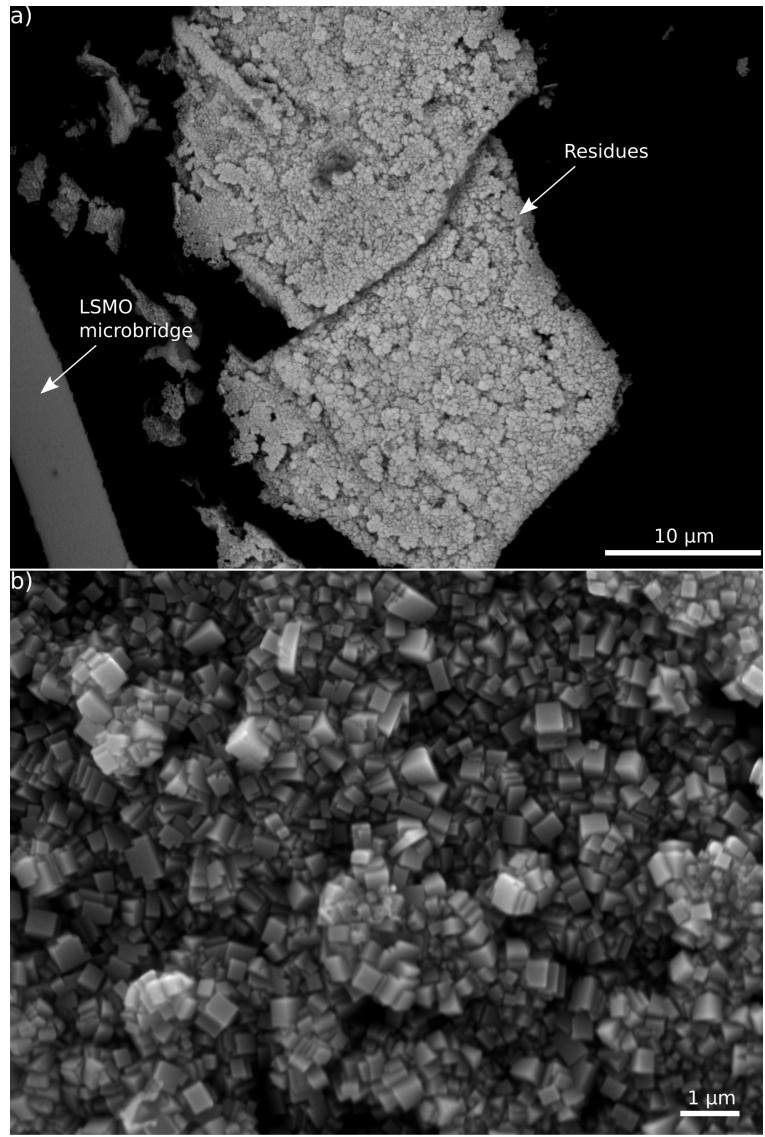

Figure S1. a) Back scattered electron image of deposits observed on the sample surface after chemical etching. b) Secondary electron image of a magnified region of the deposits. Images conditions are: Beam Accelerator Voltage (ETH) 20 kV and working distance 15 mm. These deposits are in great part gradually removed by mild agitation during the etching process. The cubic crystallites are identified as $SrF_2$.

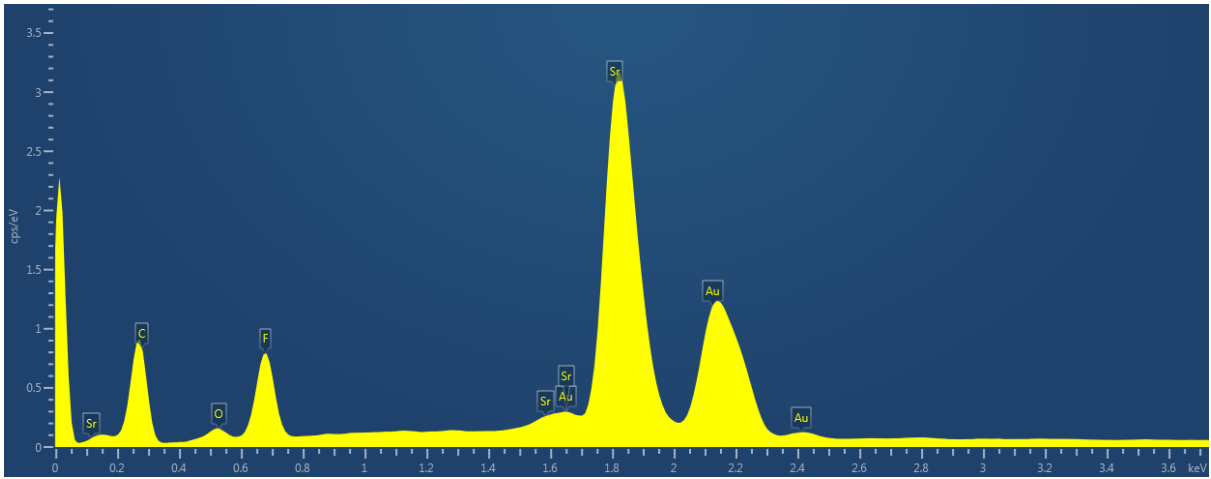

Figure S2.  Energy Dispersive X-ray spectrum acquired on the deposits of Figure S1. The sample was prepared by sticking carbon tape over the STO surface and detaching the LSMO microbridges visible in Figure S1(a) and the residues. Strontium and Fluorine are detected. The presence of carbon is justified by the metallization of the sample for SEM measurements, while gold was previously deposited on top of the sample for other characterization purpouses.

| Element | Line type | Apparent concentration | k Ratio | Wt% | Wt% sigma | Atomic % |
|---------|-----------|------------------------|---------|-----|-----------|----------|
| F | K series | 25.26 | 0.04959 | 16.67 | 0.18 | 73.15 |
| Sr | L series | 27.37 | 0.24078 | 28.22 | 0.21 | 26.85 |
| Total | | | | 44.89 | | 100 |

TABLE S1. Quantitative analysis of the deposits after the etching of $SrTiO_3$ in HF 5%

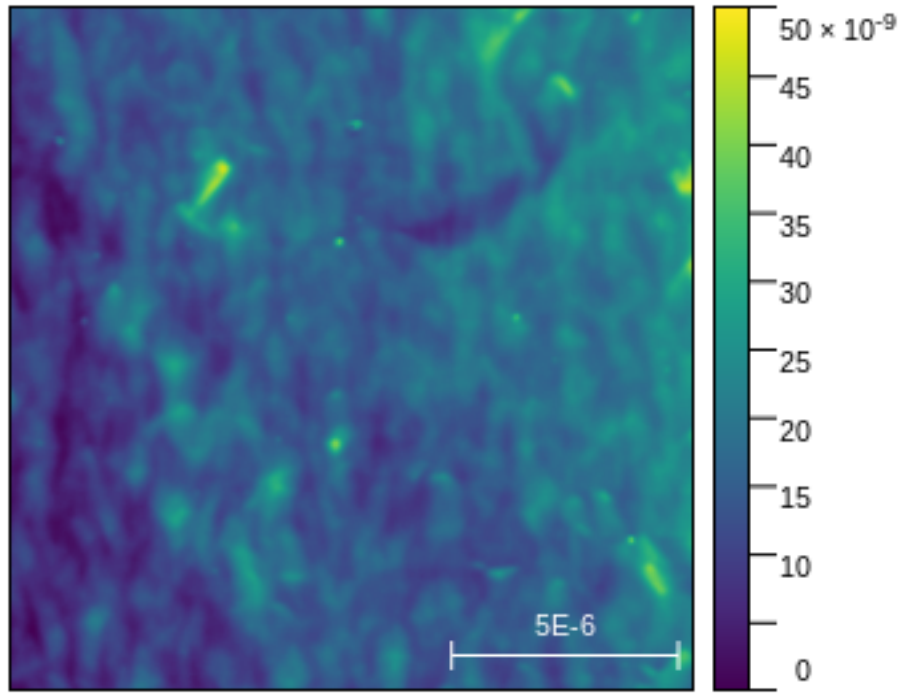

Figure S3.   Atomic force microscopy image acquired over a $15 \times 15\,\mu m^2$ area. The surface is a $SrTiO_3(110)$ sample after 60 minutes of etching in $5\,\%$ HF:$H_2O$ solution. Root-mean-square height value calculated over the whole area is $\approx 4\,nm$. This image was processed in Gwyddion by performing $2^{nd}$-order polynomial background substraction.

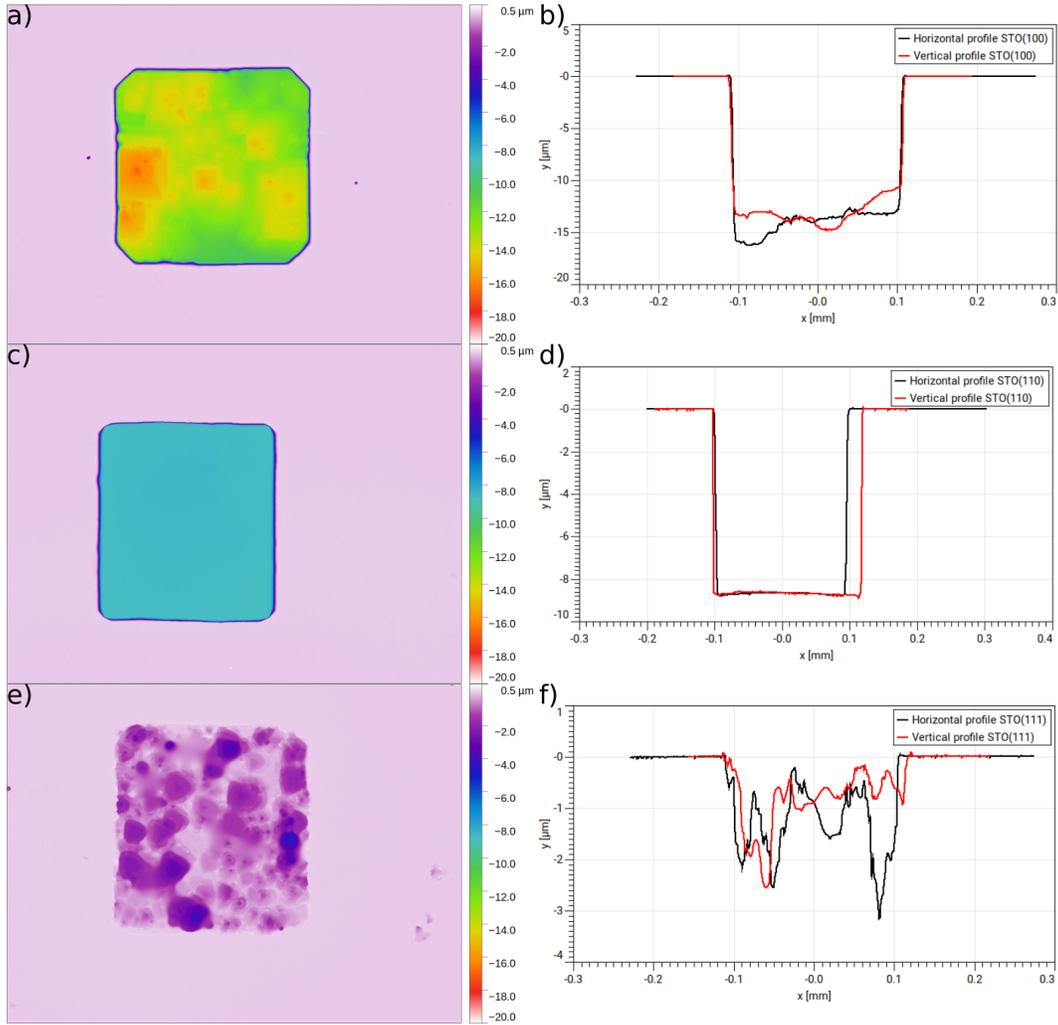

Figure S4. Depth map (left) and profiles line-cut (right) of etched SrTiO$_3$ (100) (a,b), (110) (c,d), and (111) (e,f) substrates acquired by optical profilometry. The 200×200 μm² squares were patterned using a (La,Sr)MnO$_3$ hard mask to selectively expose the SrTiO 3 surface. The samples were then put soaking for 1 h in a 5% of HF solution in distilled water. The line profiles were extracted from the central vertical and horizontal axes of the squares. Large-area etch rates and RMS (excluding the RMS for the (110) case) are reported in Table I in the main text and have been calculated from these images by selecting the largest rectangular region comprised within the squares.

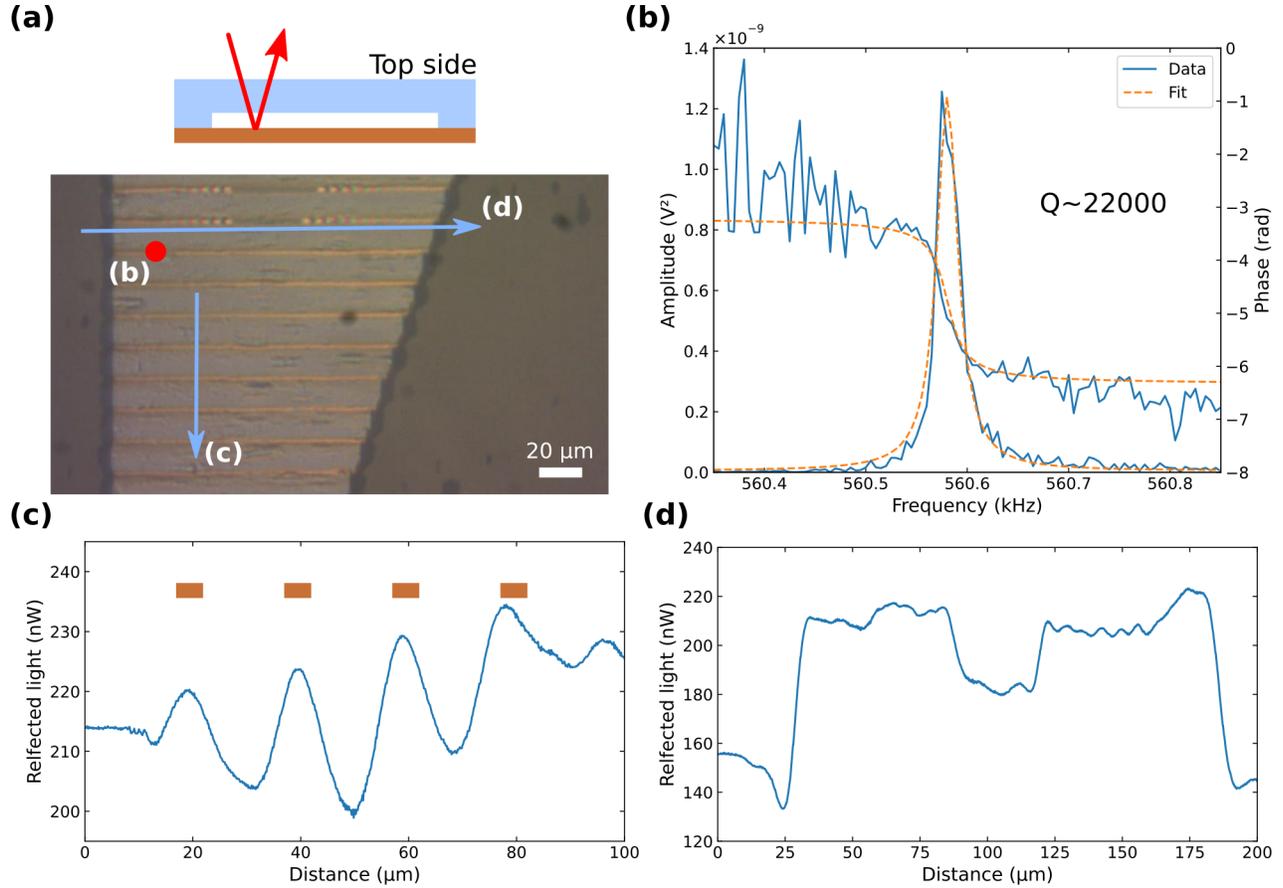

Figure S5.  (a) optical microscope image of an array of LSMO suspended microbridges acquired from the backside of a STO(110) double polished substrate. (b) Magnitude and phase (solid blue) together with Lorentzian fit (dashed orange) of the first flexural mode of one of the bridges reported in (a) measured with the laser passing through the substrate. (c)–(d) spatial dependence of the reflected light while moving the laser crossing (c) or along (d) the microbridges' length.

In Figure S5(a) we show a picture in reflected light of an array of 100 nm thick (La,Sr)MnO$_3$ (LSMO) microbridges fabricated on a 500 µm thick double-polished SrTiO$_3$(110) substrate. During the fabrication, the sample was completely immersed in 5% HF solution, with no protective layer on the backside surface. This image is taken from the backside of the substrate and the LSMO microbridges are visible through the transparent substrate itself. Thanks to the etching properties of SrTiO$_3$(110) discussed on the main text, the backside SrTiO$_3$ surface and the SrTiO$_3$ region below the freestanding LSMO

bridges remain enough flat to allow the focusing of a laser beam directly on the back side of the LSMO microbridge, as schematized in the figure. In Fig. S5(b) we show the measurement of the mechanical resonance of one of the LSMO microbridges presented in (a). The laser beam is focused on the indicated spot in the optical lever geometry and the sample is placed in vacuum ($10^{-5}$ mbar) and at $30\,°C$. A piezo crystal, glued nearby the sample and connected to a spectrum analyzer, mechanically excites the LSMO microbridges. In Fig. S5(c) we show the spatial dependence of the laser beam power during a scan of the laser beam across different microbridges, as indicated by the vertical blue arrow in (a). The microbridge positions are marked by brown rectangles and correspond to a maximum of the reflected light power. In Fig. S5(d) we perform another similar spatial mapping but during a laser scan along a single microbridge that contacts the substrate at its center. The power of the reflected beam is different in the LSMO film (edges), in the stuck LSMO microbridge (center) and in the freestanding LSMO microbridge regions (intermediate regions). These measurements show that it is possible to conceive a device architecture where the LSMO microbridges are placed in close proximity of an object and perform mechanical characterizations at the same time exploiting the transparency of the substrate of growth.



\end{document}